\documentclass[12pt,apj]{emulateapj}

\usepackage{subfigure}
\usepackage[usenames, dvipsnames]{xcolor}
\usepackage[%
  breaklinks = true,
  colorlinks = true,
  urlcolor   = blue,
  citecolor  = blue,
  linkcolor  = blue,
]{hyperref}
\usepackage{soul}

\begin{document}

\shorttitle{Non-equilibrium ionization and ambipolar diffusion}
\shortauthors{Mart\'inez-Sykora et al.}
\title{Ion-neutral interactions and non-equilibrium ionization in the solar chromosphere}

\author{Juan Mart\'inez-Sykora \altaffilmark{1,2,4,5}  
	\& Jorrit Leenaarts \altaffilmark{3}
	\& Bart De Pontieu \altaffilmark{2,4,5}	
	\& Daniel N\'obrega-Siverio \altaffilmark{4,5}
	\& Viggo H. Hansteen \altaffilmark{1,2,4,5}
	\& Mats Carlsson \altaffilmark{4,5} 
	\& Mikolaj Szydlarski \altaffilmark{4,5}}

\email{juanms@lmsal.com}
\affil{\altaffilmark{1} Bay Area Environmental Research Institute, Moffett Field, CA 94035, USA}
\affil{\altaffilmark{2} Lockheed Martin Solar and Astrophysics Laboratory, Palo Alto, CA 94304, USA}
\affil{\altaffilmark{3} Institute for Solar Physics, Department of Astronomy, Stockholm University, AlbaNova University Centre, SE-106 91, Stockholm, Sweden}
\affil{\altaffilmark{4} Institute of Theoretical Astrophysics, University of Oslo, P.O. Box 1029 Blindern, N-0315 Oslo, Norway}
\affil{\altaffilmark{5} Rosseland Center for Solar Physics, University of Oslo, P.O. Box 1029 Blindern, N-0315 Oslo, Norway}

\newcommand{\eg}{{\it e.g.,}} 
\newcommand{\myemail}{juanms@lmsal.com}
\newcommand{\komment}[1]{\texttt{#1}}
\renewcommand{\ul}{\underline}
\newcommand{\pref}{\protect\ref}
\newcommand{\soho}{{\em SOHO{}}}
\newcommand{\sdo}{{\em SDO{}}}
\newcommand{\stereo}{{\em STEREO{}}}
\newcommand{\iris}{{\em IRIS{}}}
\newcommand{\hinode}{{\em Hinode{}}}
\newcommand{\jms}[1]{\color{black}{#1}}
\newcommand{\jl}[1]{\color{blue}{#1}}
\newcommand{\bdp}[1]{\color{green}{#1}}
\newcommand{\dns}[1]{\color{yellow}{#1}}
\newcommand{\vhh}[1]{\color{orange}{#1}}
\newcommand{\mc}[1]{{\color{red}{MC: #1}}}
\newcommand{\msz}[1]{{\color{red}{M.Sz: #1}}}

\newcommand{\nongol}{{\em non\_gol\_lte}}
\newcommand{\gol}{{\em gol\_lte}}
\newcommand{\golneq}{{\em gol\_nei}}
\newcommand{\be}{\begin{equation}}
\newcommand{\ee}{\end{equation}}

\newcommand\HI{\textsc{H~I}}
\newcommand\HII{H~II}
\newcommand\HeI{He~I}
\newcommand\HeII{He~II}
\newcommand\HeIII{He~III}

\newcommand{\bea}{\begin{eqnarray}}
\newcommand{\eea}{\end{eqnarray}}

\begin{abstract}

The thermal structure of the chromosphere is regulated through a complex interaction of various heating processes, radiative cooling, and the ionization degree of the plasma. Here we study the impact on the thermal properties of the chromosphere when including the combined action of non-equilibrium ionization (NEI) of hydrogen and helium and ion-neutral interaction effects. We have performed a 2.5D radiative magnetohydrodynamic simulation including ion-neutral interaction effects by solving the generalized Ohm's law (GOL) as well as  NEI for hydrogen and helium using the Bifrost code. The GOL equation includes ambipolar diffusion and the Hall term. We compare this simulation with another simulation that computes the ionization in local thermodynamic equilibrium (LTE) including ion-neutral interaction effects. Our numerical models reveal substantial thermal differences in magneto-acoustic shocks, the wake behind the shocks, spicules, low-lying magnetic loops, and the transition region. In particular, we find that heating through ambipolar diffusion in shock wakes is substantially less efficient, while in the shock fronts themselves it is more efficient, under NEI conditions than when assuming LTE.

\end{abstract}


\section{Introduction}

Ion-neutral interaction and non-equilibrium ionization (NEI) are key physical processes in the solar chromosphere and transition region (TR). Most studies investigating these processes are focused on either one or the other process, but hitherto due to their complex physics and the associated high computational cost of modeling such systems \citep[see the review by][and references therein]{Ballester:2018fj},
not both of them combined. 

Hydrogen has an ionization/recombination timescale of $10^3$ -- $10^5$~s in the 
chromosphere \citep{Carlsson:1992kl,Carlsson:2002wl} (except in shock fronts where it is 
shorter). This is a very long time compared to typical magnetohydrodynamic (MHD) timescales, 
which are of order 10 -- 100~s. Likewise, \citet{Golding:2014fk} found that the 
ionization/recombination timescale of helium is of order $10^2$--$10^3$~s, also longer than the 
MHD timescales in the chromosphere and TR. NEI effects for heavier atoms 
occur on timescales shorter than about $10$-–$100$~s for typical density values for the transition 
region and corona \citep{2010ApJ...718..583S}. 

A major consequence of the long ionization and recombination timescales is that the solar plasma exhibits larger temperature fluctuations than what is predicted based on local thermodynamic equilibrium (LTE) ionization. When gas is suddenly heated, the extra energy is not used to ionize hydrogen or helium, but instead mainly increases the thermal energy and thus the temperature. When gas is suddenly cooled, for example by expansion in the wake of a shock front, little energy is released through recombination and the temperature drops more than in LTE. 

The ionization degree of the hydrogen (defined here as $F_\mathrm{Hion}= n_\mathrm{HI}/(n_\mathrm{HI}+n_\mathrm{HII})${\jms, where $n_\mathrm{HI}$ and $n_\mathrm{HII}$ are density number of, respectively, neutral and ionized hydrogen)} in the plasma becomes rather insensitive to temperature in NEI. In earlier 2D simulations of the solar chromosphere  $F_\mathrm{Hion}$ increases from $10^{-5}$ -- $10^{-4}$ in the upper photosphere to close to unity in the TR, while the bulk of the chromosphere has $F_\mathrm{Hion} \approx 10^{-2}$ \citep{Leenaarts:2007sf}.

Non-equilibrium ionization also removes bands of ``preferred temperatures'', corresponding to hydrogen and helium ionization temperatures in simulations that assume LTE  \citep{Leenaarts:2011qy,Golding:2016wq}. Additionally, NEI leads to differences in the emergent intensities of spectral lines, either directly because of non-equilibrium level populations \citep{Golding:2017zm}, or indirectly through the changes in the 
electron density for spectral lines that otherwise can be modelled using statistical equilibrium \citep[e.g.,][]{Wedemeyer-Bohm:2011oq,Leenaarts:2013ij}.

The chromosphere is partially ionized. The ion-neutral collision frequency is small enough for ion-neutral interaction effects to be relevant for the thermodynamics of the chromosphere \citep[e.g.][]{Vernazza:1981yq,Khomenko:2012bh,Martinez-Sykora:2017gol}. Some ion-neutral interaction effects in the chromosphere can be taken into account by relaxing the MHD constraints and using the so-called generalized Ohm's law (GOL), which includes, at least, the Hall term and ambipolar diffusion. This simplified approach improves the computational efficiency of simulations when comparing to multi-fluid models since ion-neutral interaction effects are included in a formulation using only single fluid. The approximation underlying the use of GOL is valid as long as the velocity drift between ions and neutrals is small compared to velocities from waves and flows \citep[e.g.][]{cowling1957,Braginskii:1965ul,Parker:2007lr,Ballester:2018fj}. As shown in \citet{Martinez-Sykora:2012uq}, this assumption is fulfilled in most of the chromosphere.

Ion-neutral interaction effects, i.e., the ambipolar diffusion and to some extent the Hall term, play a role in many  physical processes. First, ion-neutral interactions can cause significant damping of Alfv\'en waves, especially for high frequencies \citep{De-Pontieu:2001fj,Zaqarashvili:2012iw,Soler:2015gd}, potentially substantially reducing the energy flux that reaches the corona. Second, magnetic field could slip through the chromospheric material 
\citep{Leake2006,Arber:2007yf,Leake:2013dq,Martinez-Sykora:2016qf} and magnetic energy can be dissipated and lead to heating due to the presence of the ambipolar diffusion  \citep{Goodman:2011wq,Khomenko:2012bh,Martinez-Sykora:2017gol}.  Third, ambipolar diffusion has been shown to be a potential key process in triggering type II spicules and, associated with the formation of these spicules, in driving Alfv\'enic waves \citep{Martinez-Sykora:2017sci}. 

The simulations of \citet{Martinez-Sykora:2012uq,Martinez-Sykora:2015lq,Martinez-Sykora:2017gol} show variations of many orders of magnitude in the ionization degree, and, consequently, in the ion-neutral collision frequency within the same scale height. This is a consequence of the assumption that the ionization follows Saha-Boltzmann statistics (LTE), which predicts ionization degrees for hydrogen ranging from $\sim 10^{-10}$ for cold areas up to $\sim1$ for shocks and other hot areas in the chromosphere. Since the ambipolar diffusion scales roughly as the inverse of the ionization degree, this produces similar large variations. NEI leads to ionization degrees that are rather constant for a given plasma parcel. We thus expect large changes in the magnitude of ambipolar diffusion effects when using NEI instead of utilizing LTE ionization. The aim of this paper is to investigate the thermal properties of the chromosphere when NEI and ion-neutral effects are included at the same time.

The layout of this manuscript is as follows. We briefly describe the {\it Bifrost} code \citep{Gudiksen:2011qy}, numerical methods, the setup of the simulations and their physical processes in Section~\ref{sec:mod}. We study the impact of ion-neutral interaction effects and NEI by comparing and analyzing the following two simulations: a) one that takes into account ion-neutral interactions but also assuming LTE; and b) the other computes the ion-neutral interaction effects and the NEI of hydrogen and helium. We analyze these models in detail in Section~\ref{sec:res} and concentrate our discussion on the following topics: general aspects of the thermodynamic, ambipolar diffusion and NEI properties (Section~\ref{sec:std}); the impact of the NEI on the ambipolar diffusion (Section~\ref{sec:diff}), and their impact on the chromospheric thermodynamics (Section~\ref{sec:therm}) with special focus on the cold wake behind chromospheric magneto-acoustic shocks and cold regions within type II spicules (Section~\ref{sec:adb}), the shocks as well as the hot regions within type II spicules (Section~\ref{sec:hot}), and low-lying chromospheric and TR loops (Section~\ref{sec:high}). We end with a discussion and conclusions in Section~\ref{sec:dis}.

\section{Numerical Model}~\label{sec:mod}

\begin{figure*}
	\includegraphics[width=\textwidth]{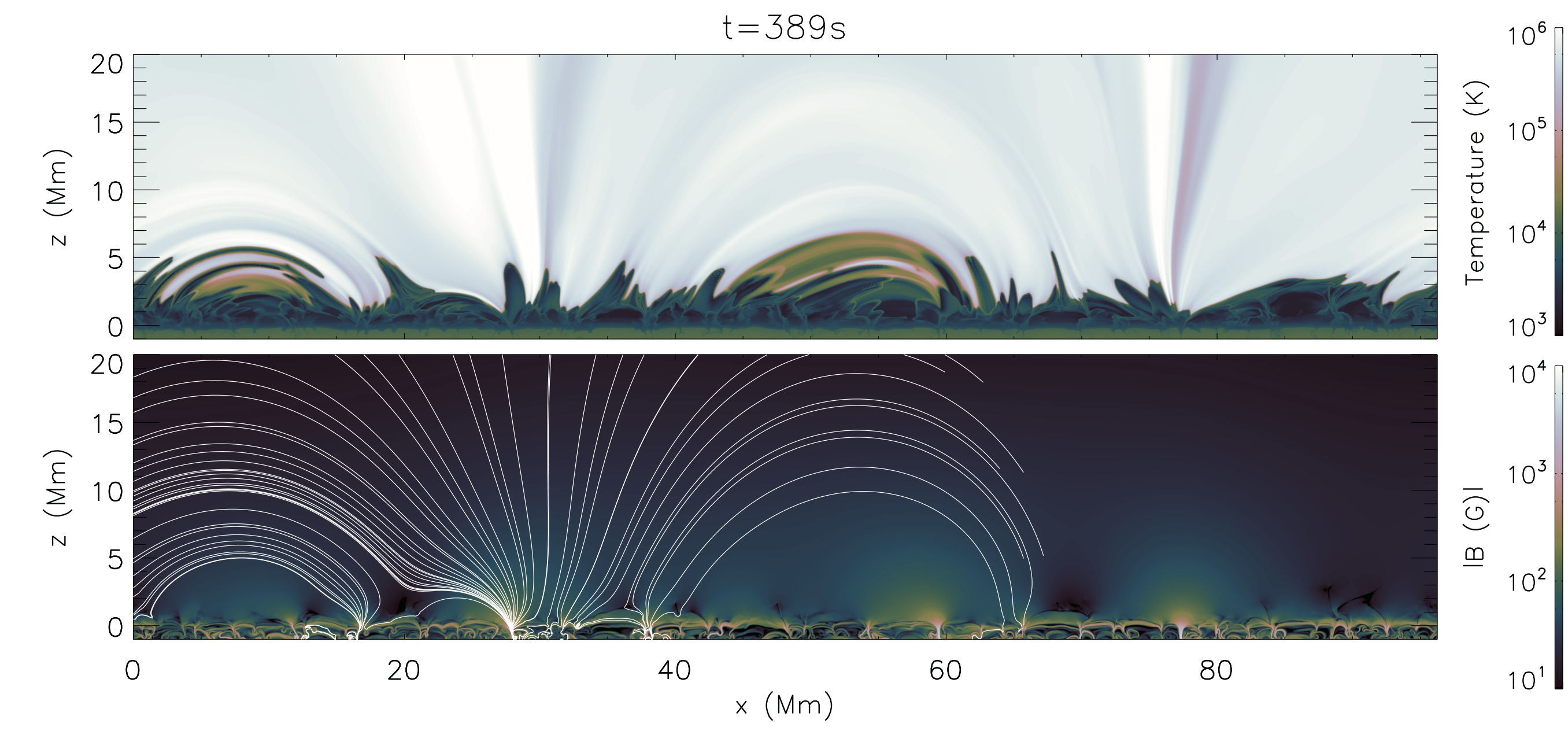}  
	\caption{\label{fig:2dtemp} Temperature (top) and magnetic field strength (bottom) in the \golneq\ simulation at $t=390$~s after switching on the NEI. Magnetic field lines (bottom) are shown on the left-hand side only in order to clearly show the magnetic field structure on the right-hand side. Note that the numerical domain spans from 3~Mm below the surface  ($z=0$~Mm) to 40 Mm above the surface, so these maps only show the lower atmosphere.  An animation of this figure is available in the online Journal showing time evolution.}
\end{figure*}

We have performed two 2.5D radiative MHD simulations using the {\it Bifrost} code. The simulations differ in their treatment of the ionization balance. These two simulations include a common set of physics packages: radiative transfer with scattering in the photosphere and lower chromosphere \citep{Skartlien2000,Hayek:2010ac}, upper chromospheric and TR radiative losses and gains following \citet{Carlsson:2012uq} recipes, optically thin radiative losses in the corona, thermal conduction along the magnetic field \citep{Gudiksen:2011qy}, and ion-neutral effects using the generalized Ohm's law \citep{Martinez-Sykora:2017gol,Nobrega-Siverio:2019tec}. In addition, an {\it ad hoc} heating term is included to ensure that the temperature does not drop far below the validity range of our equation of state ($\sim 1800$~K). 

Both simulations have the same spatial resolution, domain size, and magnetic field configuration. The numerical domain spans $90$~Mm horizontally and goes from 3~Mm below to 40 Mm above the photosphere (see Figure~\ref{fig:2dtemp} which is trimmed in order to reveal the structures of the lower atmosphere). The horizontal resolution is uniform with $14$~km separation between grid points. The vertical grid spacing is non-uniform. It is $12$~km in the photosphere, chromosphere and TR and increases in the corona and convection zone where the scale heights are larger.

The magnetic field configuration has two main opposite polarities of the size and strength (with $190$~G unsigned mean magnetic field) of a medium size plage region ($10$~Mm) separated by $45$~Mm. This leads to $\sim50$~Mm long loops connecting the two polarities. 

Concerning the boundary conditions, they are periodic in the horizontal direction and {\jms characteristic open boundaries} in the vertical direction, allowing waves and plasma to go through without reflection \citep[{\jms see the Appendix of }][{\jms for further details about the characteristic boundary implementation}]{Gudiksen:2011qy}. In addition, the bottom boundary has a constant entropy input in regions of inflow to drive and maintain the solar convective motions so the effective temperature at the photosphere of the simulations is $\sim 5780$~K. No new magnetic flux was added through the lower boundary.

Both simulations include ion-neutral interaction effects by adding two new terms in the induction equation: the Hall term and the ambipolar diffusion. The expanded induction equation is: 
 
\bea
\frac{\partial {\bf B}}{\partial t} &=& \nabla \times \left[{\bf u \times B} -  \eta {\bf J}
 - \frac{\eta_\mathrm{H}}{ |B|} {\bf J \times B} \right.  \nonumber \\
 && \left. + \frac{\eta_\mathrm{A}}{ B^2} ({\bf J \times B}) \times {\bf B}\right]\label{eq:indeq}
\eea

\noindent  where $\bf{B}$, $\bf{J}$, $\bf{u}$, $\eta$, $\eta_\mathrm{H}$, and  $\eta_\mathrm{A}$ are the magnetic field, current density, velocity field and magnetic diffusivity, Hall term coefficient, and the ambipolar diffusion coefficient. The first term on the right hand side of the induction equation is the convective term and the second term represents the Ohmic diffusion. The Ohmic diffusion is not treated in the Bifrost code since the hyper-diffusion terms typically are much larger than the Ohmic diffusion \citep{Martinez-Sykora:2017gol}. The third term, the Hall term, does not lead to energy dissipation, but the ambipolar diffusion term (the last one) leads to additional heating owing to dissipation of currents that are perpendicular to the magnetic field:
\be
Q_\mathrm{A} =  \eta_\mathrm{A} \mathbf{J}^2_\perp,
\ee
with $Q_\mathrm{A}$ the ambipolar heating per unit volume and $ \mathbf{J}_\perp$ the current density perpendicular to the magnetic field.

The Hall term depends on the electron number density while the ambipolar diffusion depends on the ion-neutral collision frequency and ionization fraction: 

\begin{eqnarray}
\eta_\mathrm{H} & = & \frac{ |{\bf B}|}{q_e n_e}, \\
\eta_\mathrm{A} & = & \frac{(\rho_n/\rho)^2|\bf{B}|^2}{\sum_a \sum_{a'} \rho_{a0} \frac{m_{a'\hat{I}}}{m_{a'\hat{I}} + m_{a0}}\nu_{a0a'\hat{I}}}; \label{eq:diff}
\end{eqnarray}

\noindent where for clarity and consistency we used the same nomenclature as in \citet{Ballester:2018fj}, i.e., species and the ionization states are referred with the sub-index $a$, and $I$, respectively, i.e., $I=0$ denotes neutrals and $\hat{I} = I \geq 1$ ions. $\rho_\alpha$, $m_\alpha$, and  $\nu_{\alpha\beta}$ for any $\alpha \in [a0,a\hat{I}]$ or $\beta \in [a'0,a'\hat{I}]$ are the mass density, atomic mass and ion-neutral collision frequency between various species. $\rho$ and $\rho_n$ are the total mass density and total neutral mass density ($\rho_n=\Sigma_a \rho_{a0}$). Finally, $n_{e}$, and $q_e$ are the electron number density and the electron charge. Note that $n_{a'\hat{I}} \nu_{a'\hat{I}a0} = n_{a0} \nu_{a0a'\hat{I}}$ as required by momentum conservation. Equation~\ref{eq:diff} assumes that all ionized species moves together. The ion-neutral collision frequency is:
\be
\nu_{a0a'\hat{I}} = n_{a\hat{I}} \, \sigma_{a0a'\hat{I}} \sqrt{\frac{8\,k_\mathrm{B}\,T}{\pi \mu_{a0a'\hat{I}}}},
\ee
with $T$ the temperature, $\sigma_{a0a'\hat{I}}$ the cross section between an ionized species and neutral species, and $k_\mathrm{B}$ is the Boltzmann constant \citep{Mitchner:1973,Vranjes:2008uq}. The quantity $\mu_{a0a'\hat{I}}$ is the reduced mass $\mu_{a0a'\hat{I}} = m_{a0} m_{a\hat{I}}/(m_{a0} + m_{a\hat{I}})$. The ion-neutral cross sections used are described in \citet{Vranjes:2013ve}. For further details about the ambipolar diffusion implementation see the paper by \citet{Martinez-Sykora:2017gol,Nobrega-Siverio:2019tec}.

In order to have a rough understanding of how the ambipolar diffusion coefficient varies with  ionization  fraction, let us consider a gas consisting of only hydrogen and assuming a cross section independent of the thermal properties of the plasma, the ambipolar diffusion has the following proportionality:
\be
\eta_\mathrm{A} \sim  \frac{|{\bf B}|^2}{n_\mathrm{H}^2 \sqrt{T}} \frac{1-F_\mathrm{Hion}}{F_\mathrm{Hion}}, \label{eq:amb_proportionality}
\ee
with $n_\mathrm{H}=n_\mathrm{H I} + n_\mathrm{H II}$. As long as $F_\mathrm{Hion} < 0.1$, as is the case in most of the chromosphere, then the dependence on the ionization degree is approximately given by
\be
\eta_\mathrm{A} \propto  \frac{1}{F_\mathrm{Hion}}.\label{eq:proxamb}
\ee
This has been done assuming only hydrogen, but note that one can conclude the same for the total ionization fraction ($F_\mathrm{ion}$). Taking into account the assumptions to reach equation~\ref{eq:proxamb}, it allows us to roughly estimate and understand the dependence of the ambipolar diffusion on the ionization fraction. 

Our first numerical simulation (\gol) includes the Hall term and the ambipolar diffusion in LTE. This simulation is the one called ``GOL" in \citet{Martinez-Sykora:2017gol,Martinez-Sykora:2017sci}. 
Our second numerical simulation (\golneq) also includes the Hall term and ambipolar diffusion just as in \gol, but instead of assuming LTE for the ionization balance it computes the ionization balance in non-equilibrium for hydrogen and helium. This is done through solving equations for charge and energy conservation together with advection and time-dependent rate equations for each level of a six-level hydrogen atom and a three-level helium atom  \citet{Leenaarts:2007sf} and \citet{Golding:2016wq}. Ionization is still assumed to follow LTE for all other species. {\jms For the non-equilibrium ionization case, the ionization rate equations for hydrogen and helium are solved:

\begin{eqnarray}
\frac{\partial n_{aIE}}{\partial t} + \nabla \cdot (n_{aIE} {\bf u}) &=& \sum_{I'E' \neq IE}^{n_{al}} n_{aI'E'} P_{aI'E'IE} -  \nonumber \\ 
&& n_{aIE} \sum_{I'E' \neq IE}^{n_{al}} P_{aIEI'E'} \label{eq:nei}
\end{eqnarray}

\noindent where $n_{al}$ is the number of levels and $P_{aIEI'E'}$ the transition rate coefficient between levels $IE$ and $I'E'$. These equations are solved using operator splitting where the continuity part (left hand side of the Equation \ref{eq:nei}) is,  like for the MHD equations, stepped forward in time using the modified explicit third-order predictor-corrector procedure \citep{hyman1979} allowing variations in time. After the predictor step, the rate part of the equations (right hand side of Equation \ref{eq:nei}) is calculated and the fundamental variables and hydrogen and helium populations are updated after a full time-step \citep{Leenaarts:2007sf,Golding:2016wq}. As mentioned, the energy and charge are conserved and are detailed in \citep{Golding:2016wq}. H$_2$ molecules are also treated in non-equilibrium \citep[see][for details]{Leenaarts:2011qy}, note that  only three body reactions have been used.} The non-equilibrium ionization degrees and electron densities are used in the computation of $\eta_\mathrm{H}$ and $\eta_\mathrm{A}$.

We refer to \citet{Martinez-Sykora:2017gol} for further details on the setup of the \gol\ simulation. {\jms In order to mitigate the numerical expenses of these simulations due to the NEI and/or GOL, we} first ran the simulation {\jms without} ambipolar diffusion and the Hall term for 35 minutes. Once transients from the initial conditions disappeared, we carried out the \gol\ simulation. Finally, when this simulation also reached a steady situation (after 15 minutes), we ran the \golneq\ simulation. All simulations were run for another 12 minutes, at least, after transients from the initial setup for each case have disappeared, but we note that the simulations are not started from the same initial conditions and thus only can be compared in a statistical sense.{\jms These transients disappear after a ``first" shock or wave goes through the chromosphere \citep[e.g. see movie of][]{Leenaarts:2007sf}. }

\section{Results}~\label{sec:res}

\citet{Martinez-Sykora:2017gol,Martinez-Sykora:2017sci,Martinez-Sykora:2018gf,De-Pontieu:2017pcd} and \citet{Khomenko:2018rm} have, using radiative MHD numerical simulations, revealed the importance of ion-neutral interaction effects in the solar chromosphere. 
Likewise, \citet{Carlsson:1992kl,Carlsson+Stein1994,Carlsson:1997ys} and \citet{Carlsson:2002wl} pointed out the importance of long timescales for NEI in the chromosphere. This was confirmed using 2D and 3D models by \citet{Leenaarts+etal2007,Leenaarts:2011qy} and \citet{Golding:2016wq,Golding:2017zm}. Here, we will show that these effects must be considered together since the thermodynamics differ substantially between simulations that assume either LTE and/or a fully ionized atmosphere.

The differences between the \gol\ simulation and a case without ambipolar diffusion have already been described in detail in \cite{Martinez-Sykora:2017gol}. We briefly summarize their main findings here: 1) numerical simulations without ambipolar diffusion do not generate type~II spicules, 2) cold bubbles that form behind the magneto-acoustic shocks due to the expansion (rarefraction) of the plasma are rather hot in the \gol\ simulation because of heating through ambipolar diffusion. 

\begin{figure*}
	\includegraphics[width=0.99\textwidth]{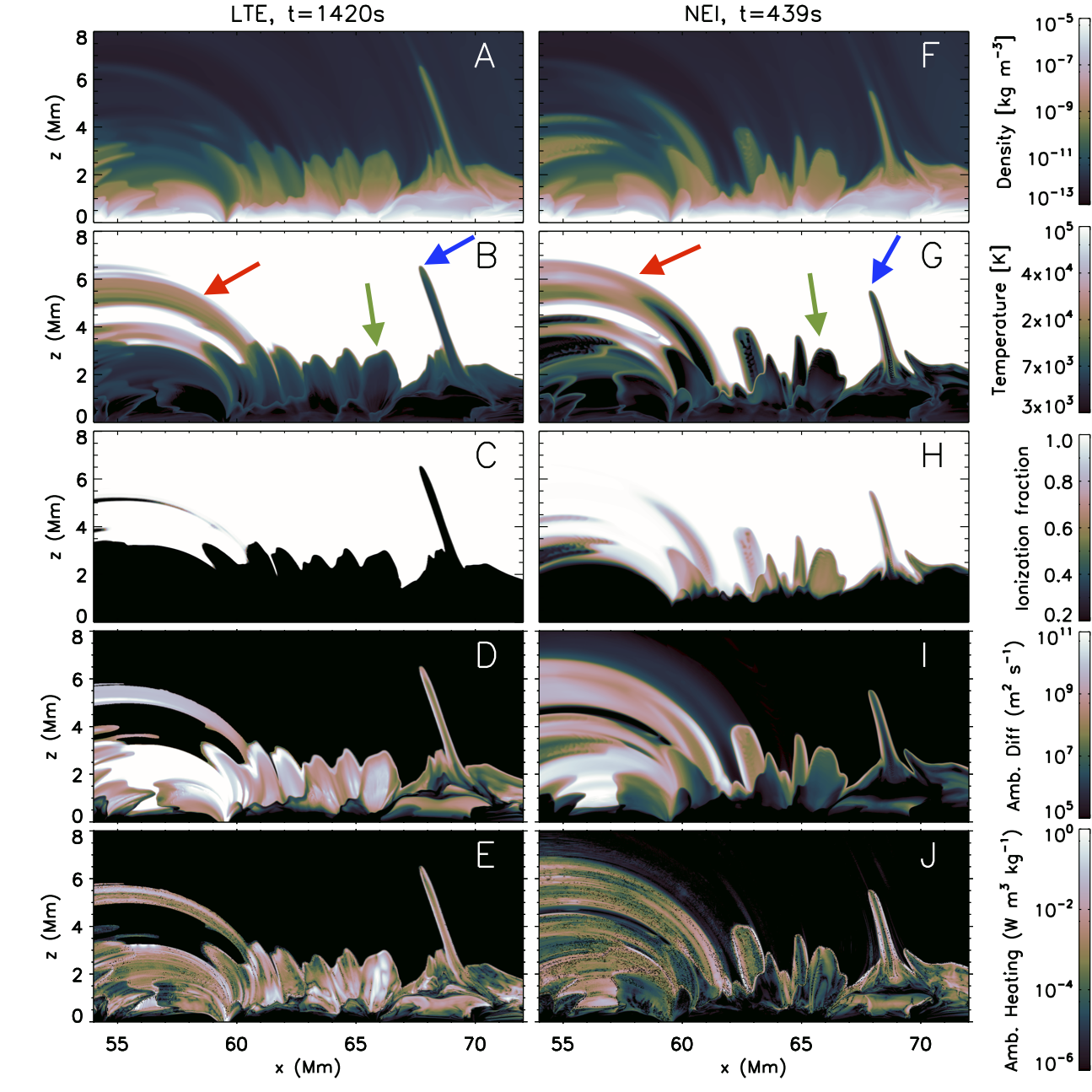}  
	\caption{\label{fig:denstg_zoom} The thermodynamics in the chromosphere differs substantially between each model. Density (A, F), temperature (B, G), ionization fraction (C, H), ambipolar diffusion coefficient (D, I) and ambipolar heating (E, J) for the \gol\ (left), and \golneq\ (right) simulations.  The blue arrows point to a type~II spicule in the simulations. The green arrows point to expanding bubbles of chromospheric gas following the passage of a shock wave and the red arrows point to  low-lying loops. An animation of this figure is available in the online Journal showing time evolution. }
\end{figure*}

Non-equilibrium ionization influences the effects of ambipolar diffusion, and the combination of these two processes (\golneq) leads to different thermal structures than what is found when neither or only one of these processes is considered. Figure~\ref{fig:denstg_zoom} zooms in on the chromosphere for each simulation: As in the \gol\ simulation, in \golneq\ there are type~II spicules (blue arrow in panel G) and low-lying loops (red arrow), but these spicules and loops have greater variations in temperature than those found in \gol . {\jms Note that, within $z=[1,6]$~Mm, loops, shocks and spicules have darker  ($\sim 2\, 10^3$~K) and lighter colors ($\sim 4\, 10^4$~K) in panel G than in panel B, where plasma is mostly accumulated around the temperature of ionization of hydrogen in LTE, i.e.,  $\sim 6 \, 10^{3}$~K (see also Section~\ref{sec:therm})}. The cold bubbles behind magneto-acoustic shocks (green arrows), mostly in the upper chromosphere, reach very low temperatures while magneto-acoustic shocks are {\jms hotter} in the \golneq\ simulation (panel G). The thermal differences between \gol\ and \golneq\ come from the ionization fraction (panels C and H) and the ambipolar diffusion (panels D and I). Any heating or cooling in NEI will increase or decrease the plasma temperature instead of ionizing or recombining hydrogen or helium in the medium. In addition, since the ionization degree is drastically different when assuming LTE (panel C) as compared to assuming NEI (panel H), the effects of ambipolar diffusion also differ (panels D and I).  As a result, the location where ambipolar heating (panels E and J) occurs has changed, being large in some locations within type II spicules, low-lying loops and in the shocks, but drastically reduced in cold bubbles and regions with strong expansion.

\subsection{Statistical properties}~\label{sec:std}

We now proceed with making a statistical analysis of the ionization fraction and ambipolar diffusion (Section~\ref{sec:stdamb}) and of the thermodynamics within the chromosphere and transition region  (Section~\ref{sec:stdtg}). 

\subsubsection{On the ionization fraction and ambipolar diffusion}~\label{sec:stdamb}

\begin{figure}
    \includegraphics[width=0.49\textwidth]{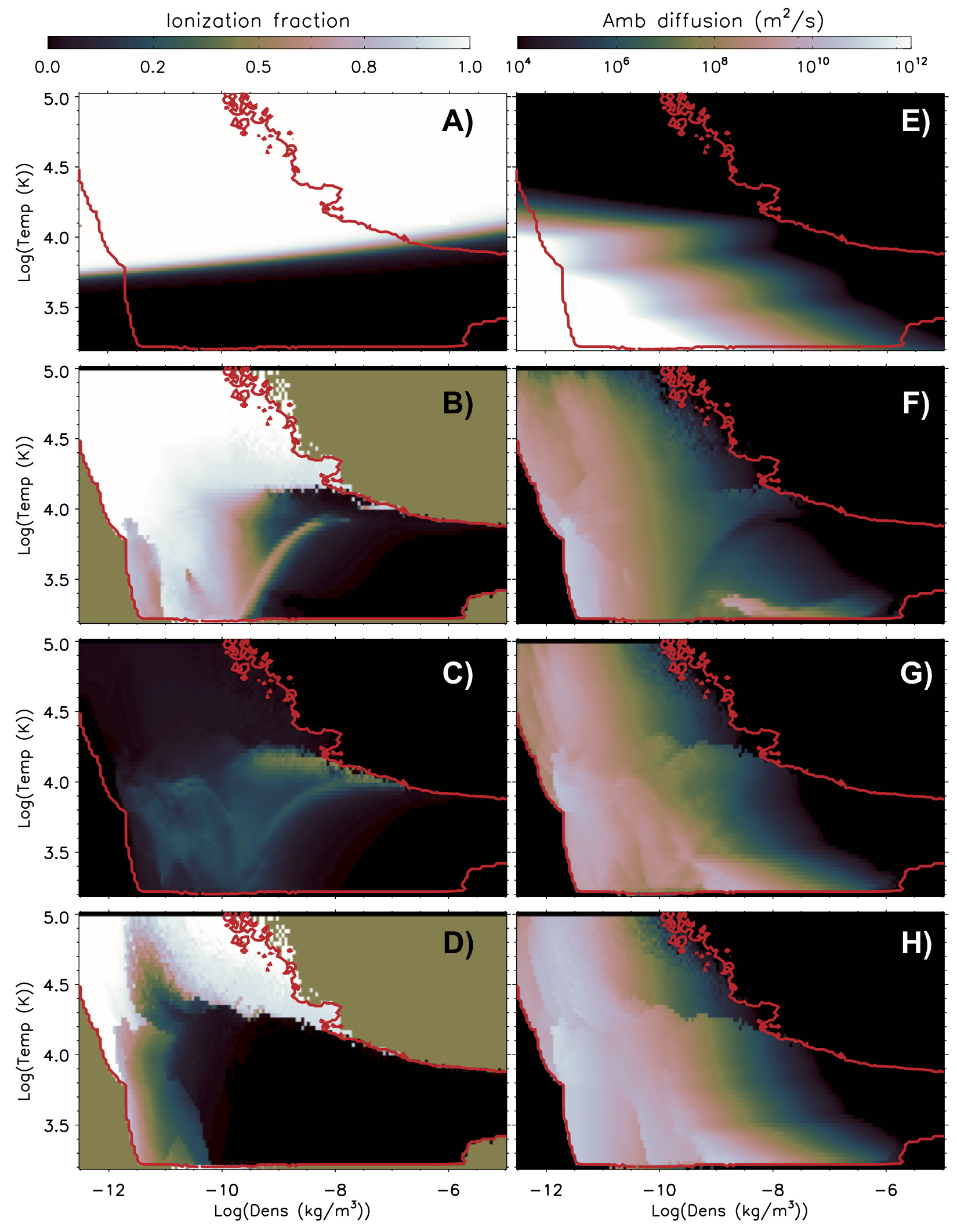}
	\caption{\label{fig:etapdf} Distribution of the ionization fraction (left) and ambipolar diffusion coefficient (right) as a function of mass density and temperature assuming a constant magnetic field strength of 30~G computed from a 12 minute period of the simulation. Panel A and E: assuming LTE. Panels B--D and F--H: assuming NEI, as computed from simulation \golneq. For this simulation the ionization fraction and ambipolar diffusion are no longer uniquely determined by the temperature and mass density. We show the median of $F_\mathrm{ion}$ (B) and $\eta_\mathrm{A}$ (F), standard deviation of $F_\mathrm{ion}$ (C) and $\log \eta_\mathrm{A}$ (G), and the minimum of  $F_\mathrm{ion}$ (D) and the maximum of $\eta_\mathrm{A}$ (H). The red contours delineate the extrema in density and temperature for the \golneq\ simulation.}
\end{figure}

When assuming LTE, both the ionization fraction ($F_\mathrm{ion}$) and ambipolar diffusion coefficient ($\eta_\mathrm{A}$) for a given magnetic field strength are uniquely defined by the temperature and density. This is because the ionisation degree in LTE is set by Saha-Boltzmann statistics only. This is shown in panel A of Figure~\ref{fig:etapdf} for $F_\mathrm{ion}$ and in panel E for $\eta_\mathrm{A}$ for a magnetic field strength of 30~G. However, in NEI, $\eta_\mathrm{A}$, {\jms calculated from \golneq ,} can vary by many orders of magnitude for a given temperature and density  (panels F-H, up to $\sim10$ orders of magnitude), because in this case the ionization fraction also depends on the history of the plasma (panels B-D).  

The ionization fraction and ambipolar diffusion has very different values when assuming LTE or NEI.\newline
In LTE  {\jms (calculated from the LTE equation of state)}, 
\begin{itemize}
    \item  $F_\mathrm{ion}$ is small (highly neutral) below $\log T = [3.7-4]$~K, and large (highly ionized) above these temperatures (see panel A in Figure~\ref{fig:etapdf}). 
    \item $\eta_\mathrm{A}$ becomes extremely large (up to $\sim 10^{16}$~m$^{2}$~s$^{-1}$) in the cold, low density areas in the chromosphere and negligible at high temperatures in upper chromosphere ($T>10^4$~K) \citep[see panel E in Figure~\ref{fig:etapdf} and][]{Martinez-Sykora:2017gol}.
\end{itemize}
In NEI {\jms (calculated from \golneq)} , 
\begin{itemize}
    \item the median of the ionization fraction (panel B) is close to fully ionized and the ambipolar diffusion coefficient (panel F) is at least five orders of magnitude smaller in regions with low densities ($\rho < 10^{-8}$~kg~m$^{-3}$) and low temperatures ($T < 10^{4}$~K) than what is obtained using LTE.
    \item The ambipolar diffusion coefficient is much larger at high temperatures in the upper chromosphere ($T>10^4$~K) (panels F and H) as compared to the LTE case because plasma, in NEI, is not fully ionized in these regions. 
\end{itemize}
Consequently we conclude that in order to properly calculate ambipolar diffusion in the solar atmosphere, a proper treatment of the ionization fraction is required. As mentioned above, as a zero order approximation, the ionization fraction and ambipolar diffusion coefficient are related by Equation~\ref{eq:proxamb}. One can consider the validity of this relation by comparing the left and right column of Figure~\ref{fig:etapdf}. However, this relation is not perfect since the ambipolar diffusion coefficient also depends on the number density and the temperature.

\begin{figure}
    \includegraphics[width=0.49\textwidth]{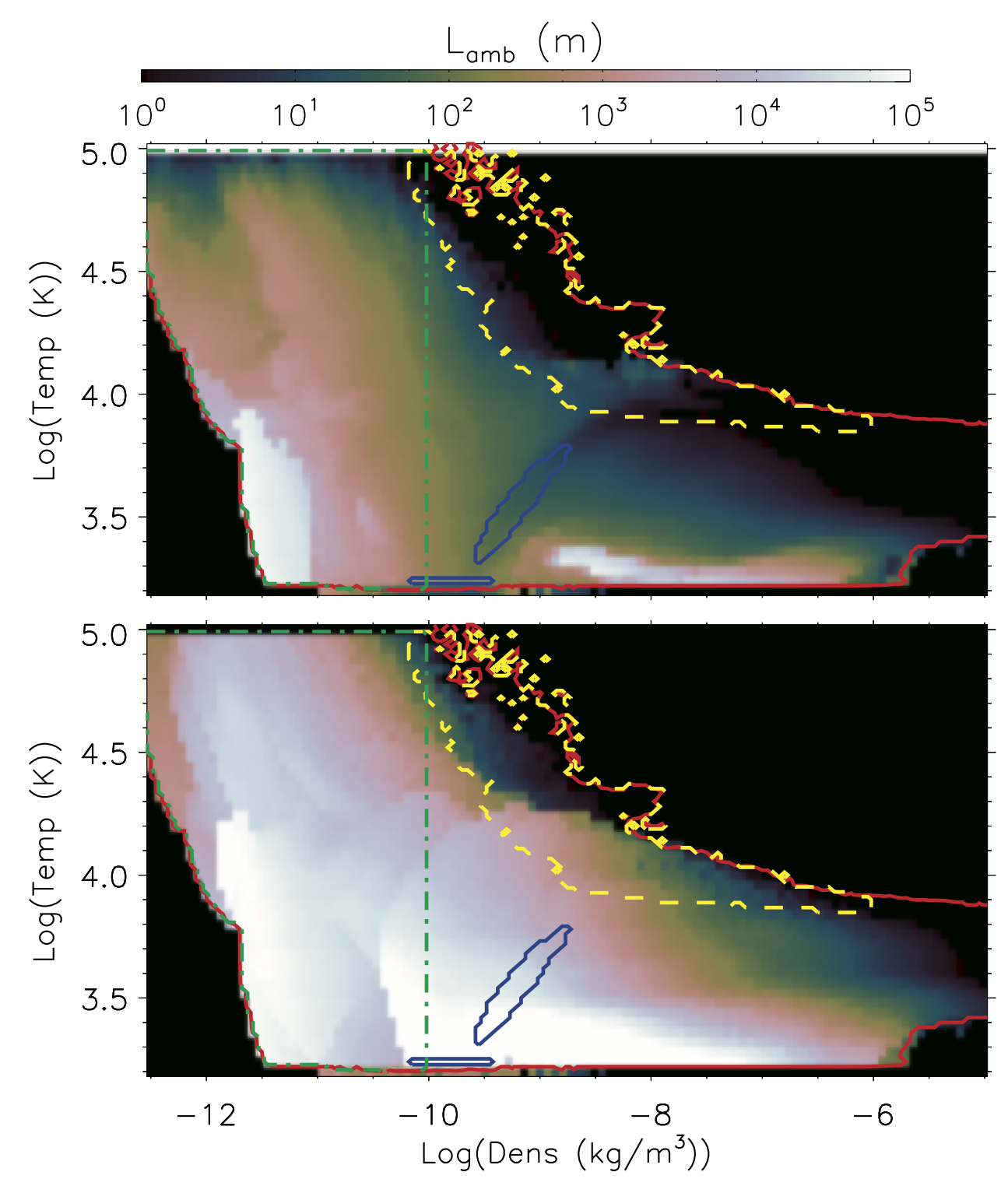}
	\caption{\label{fig:lensclpdf}  
	The ambipolar diffusion in \golneq\ acts on currents perpendicular to the magnetic field with length scales of a few tens of kilometers. The distribution of the median (top) and maximum (bottom) of $L_\mathrm{amb}$ are shown for the \golneq\ simulation as function of temperature and density integrated over 12 minutes. The red contours delineate the temperature and density range of the entire simulated atmosphere. Solid blue, dashed yellow, and dash-dotted green contours correspond to populations 1, 2 and 3 described in Figure~\ref{fig:pdf} and Sections~\ref{sec:adb}-\ref{sec:high}. }
\end{figure}

One method for estimating the importance of ambipolar diffusion is by dividing it with the Alfv\'en speed ($u_\mathrm{a}$): $L_\mathrm{amb} = \eta_\mathrm{A}/u_\mathrm{a}$. The quantity $L_\mathrm{amb}$ is the length scale on which ambipolar diffusion acts. Alfv\'en waves with wavelength similar to $L_\mathrm{amb}$, or currents perpendicular to the magnetic field ($J_\perp$) with characteristic spatial scales equal or smaller than $L_\mathrm{amb}$ will be efficiently diffused. 
Figure~\ref{fig:lensclpdf} shows the median and maximum values of $L_\mathrm{amb}$ in simulation \golneq. Ambipolar diffusion can dissipate $J_\perp$ on length scales as large as some tens of kilometers (up to a megameter) in extended regions in chromospheric and TR plasma. The median of $L_\mathrm{amb}$ is above 10~km ($10-10^3$~km) for densities $10^{-9} < \rho < 10^{-7}$~kg~m$^{-3}$ and low temperatures ($T\leq 3 \times 10^3$~K) as well as at low densities ($\rho < 10^{-10}$~kg~m$^{-3}$) and temperatures $T\leq 10^4$~K. 
In addition, some plasma reaches large ambipolar diffusivities at densities $\rho < 4\times 10^{-10}$~kg~m$^{-3}$ and up to TR temperatures (see pink-white colors in the bottom panel of Figure~\ref{fig:lensclpdf}). Note that the spatial grid spacing is $\sim10$~km in this simulation, so anything smaller will not be resolved in the numerical simulation. Most of the currents in the simulation are highly confined, i.e., a few tens of kilometers \citep[e.g.,][]{Hansteen:2015qv,Nobrega-Siverio:2016qf,Martinez-Sykora:2017rb}. 

\subsubsection{On the thermal structure in the chromosphere}~\label{sec:stdtg}

\begin{figure}
	\includegraphics[width=0.49\textwidth]{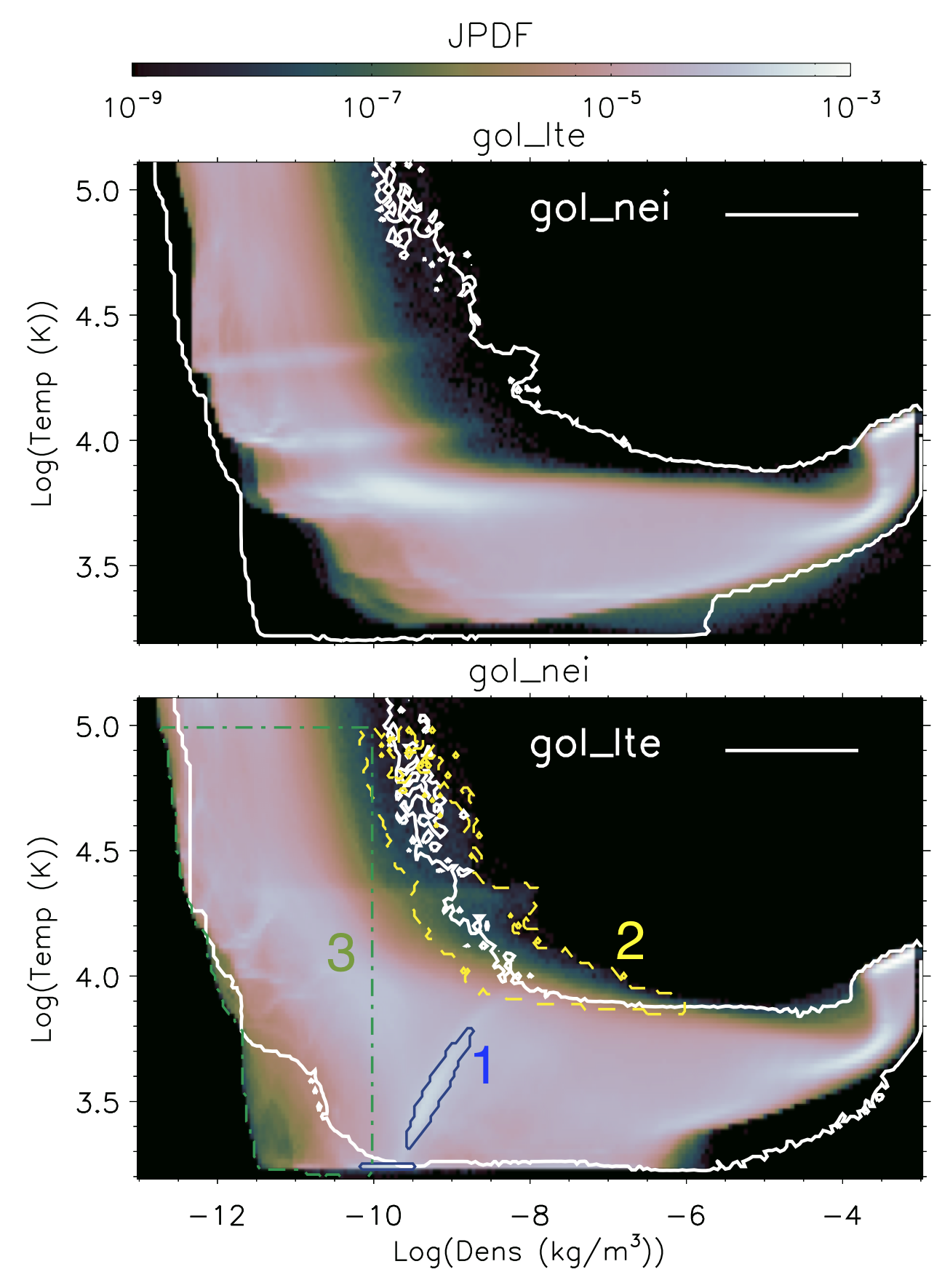}  
	\caption{\label{fig:pdf} Statistical thermal properties show differences between the models revealing large impact of ion-neutral interaction effects and NEI. Joint Probability Distribution Function (JPDF) of temperature and density for \gol\ (upper panel) and \golneq\ (lower panel) simulations, each computed from a time series of 12 minutes of solar time. The white contours correspond to the temperature and density regime of the whole simulation at JPDF$=5\times 10^{-5}$ for the other simulation (see labels). 
	Solid blue, yellow dashed and green dash-dotted contours are selected regions 1, 2 and 3 described in detail in Sections~\ref{sec:adb}-\ref{sec:high}.}
\end{figure}

Figure~\ref{fig:pdf} shows Joint Probability Distribution  Functions (JPDF) of the density and temperature in each simulation computed during a period of 12 minutes of solar time. 
Below we discuss the statistical differences in the thermal properties between the two simulations. 

\begin{itemize}
\item  Simulation \gol\ (top panel of Figure~\ref{fig:pdf}) shows three horizontal pink-white stripes at $\log T =  10^{3.8}$, $10^4$, and $10^{4.3}$. These temperatures mark the transition from \HI\ to \HII, \HeI\ to \HeII, and \HeII\ to \HeIII\ with an LTE ionization balance. A large energy gain or loss is needed for gas to pass these temperatures, and consequently gas temperatures tend to cluster there. In the \golneq\ simulation these stripes are not present owing to the slow ionization/recombination rates. This process is explained in detail in \citet{Leenaarts:2011qy} and \citet{Golding:2016wq}. In \golneq, the low-density upper chromosphere has a large range of temperatures. Note that, for instance, around $\rho \approx 10^{-10}$ kg~m$^{-3}$ the pink-white patch extends over an order of magnitude in temperature due to the presence of NEI and this is where ambipolar diffusion can play a big role.  The plasma in this temperature/density range corresponds mainly to the type II spicules and low-lying loops described in Sections~\ref{sec:hot}-\ref{sec:high}.

\item \cite{Martinez-Sykora:2017gol} showed that the gas located in the wake of shocks (cold chromospheric bubbles) is efficiently heated in the \gol\ model. The larger temperatures in the wake of the shocks in the \gol\ simulation (top panel of Figure~\ref{fig:pdf}) result from heating through large ambipolar diffusion. The bubbles are cold and have low densities owing to the expansion,  this plasma state leads to a large ambipolar diffusion (Equation~\ref{eq:amb_proportionality}) and the ambipolar diffusion dissipates and/or advects almost all the currents that are perpendicular to the magnetic field out of the cold bubbles. 
 
When hydrogen and helium are treated in NEI (\golneq, bottom panel of Figure~\ref{fig:pdf}), extended regions of the cold chromospheric bubbles reach the minimum temperature set by the {\it ad hoc} heating term. Neither recombination nor the formation of H$_2$ molecules occurs fast enough to counter the cooling caused by adiabatic expansion. Ambipolar diffusion is inefficient, in contrast to the \gol\ model: the ionization degree remains on the order of $10^{-2}$ in NEI despite the low temperatures and the ambipolar diffusion is therefore much weaker than in LTE where the ionization degree can be as low as $10^{-10}$. The \golneq\ simulation does not reach even lower temperatures because of the {\it ad hoc} heating term in the model preventing temperatures from dropping below $\sim 1800$~K. 

\item The \golneq\ simulation has plasma with $-10<\log \rho<-6$ with temperatures higher than reached in the \gol\ model. This is indicated with the yellow contour labeled 2 in the bottom panel of Figure~\ref{fig:pdf}. We analyze this population in Section~\ref{sec:hot}. 

\item The transition region is less sharp in the \golneq\ simulation than in the \gol . The JPDF shows slightly greater values within the TR in the \golneq\ simulation (bottom panel).  In addition, the density variation is largest in the \golneq\ simulation at low TR temperatures ($4.0 \le \log (T) \le 5.0$). These thermal properties are due to the thermal evolution of type II spicules and low-lying loops  resulting from NEI and ambipolar diffusion. See the detailed description of type II spicules and low-lying loops in Sections~\ref{sec:hot} and~\ref{sec:high}. 

\item Finally, in \gol , a few dense ($\rho > 10^{-6}$ kg~m$^{-3}$) chromospheric plasma elements ($JPDF\approx10^{-7}$) reach lower temperatures than in the \golneq\ simulation. The dense chromosphere with low temperatures in  the \gol\ simulation corresponds to a few regions which have accumulated enough magnetic flux in the photosphere to emerge. In such locations, the ambipolar diffusion allows the magnetic flux to expand, carrying dense plasma and producing rather large, cold bubbles  \citep{paper1,Tortosa2009,Ortiz:2014wj}. We do not see enough of such regions of accumulated magnetic flux at the photospheric heights in the \golneq\ simulation to produce similar events. 
\end{itemize}

\subsection{Impact of non-equilibrium ionization on the ambipolar diffusion}~\label{sec:diff}

\begin{figure*}
	\includegraphics[width=0.99\textwidth]{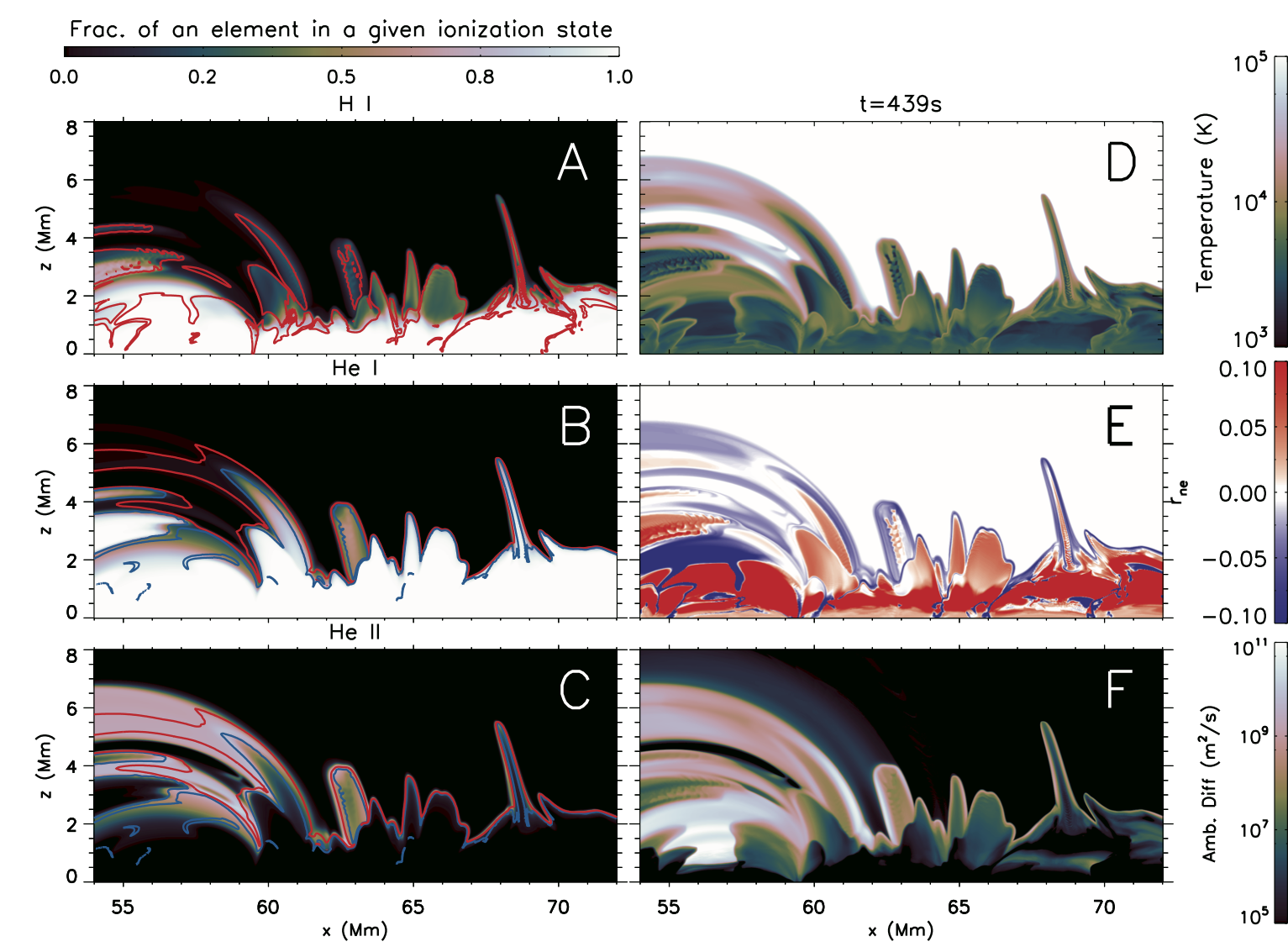}  
	\caption{\label{fig:ionf} H and He are out of equilibrium in the chromosphere which plays a role on the ambipolar diffusion coefficient. {\jms For this figure we used only the \golneq\ simulation}. Panel A-C: Fraction of all atoms as H I, He I, and He II for the \golneq\ simulation. The red contour in panel A shows where $T=6 \times 10^3$~K, roughly where H~I ionizes in LTE. The blue and red contours in panels B and C show $T=10^4$~K and $T=2.2\times 10^4$~K, the temperatures where He~I and He~II start to ionize in LTE. Panel D: temperature in \golneq. Panel E: The difference in electron density in NEI and LTE, as in Equation~\ref{eq:rne}. This reveals regions where the ionization is out of equilibrium. Panel F: the ambipolar diffusion coefficient in \golneq.  An animation of this figure is available in the online Journal showing time evolution.}
\end{figure*}

The large disparity between the amplitude of ambipolar diffusion in LTE and in NEI is mainly due to the ionization fraction (see Equation~\ref{eq:diff}), with a smaller contribution from temperature differences. {\jms The results of this section are related to the \golneq\ simulation unless otherwise it is specified. }
Figure~\ref{fig:ionf} and the associated animation show the fraction of H I, He I, and  He II, as well as temperature, and the coefficient of the ambipolar diffusion. Panel E compares the electron number density assuming NEI ($n_{\mathrm{e}_\mathrm{NEI}}$) with the electron number density computed in LTE ($n_{\mathrm{e}_\mathrm{LTE}}$) following the expression: 
\begin{eqnarray}
r_\mathrm{ne} = \frac{n_{\mathrm{e}_\mathrm{NEI}}-n_{\mathrm{e}_\mathrm{LTE}}}{n_{\mathrm{e}_\mathrm{NEI}}+n_{\mathrm{e}_\mathrm{LTE}}}. \label{eq:rne}
\end{eqnarray}
 
Blue in panel~E means that the electron number density, and therefore the ion number density, is smaller in NEI than in LTE, and red means the opposite. Under NEI conditions, the ionization fraction in the chromosphere departs from their corresponding ionization temperatures in LTE (the overlaying contours).

\begin{figure}
	\includegraphics[width=0.5\textwidth]{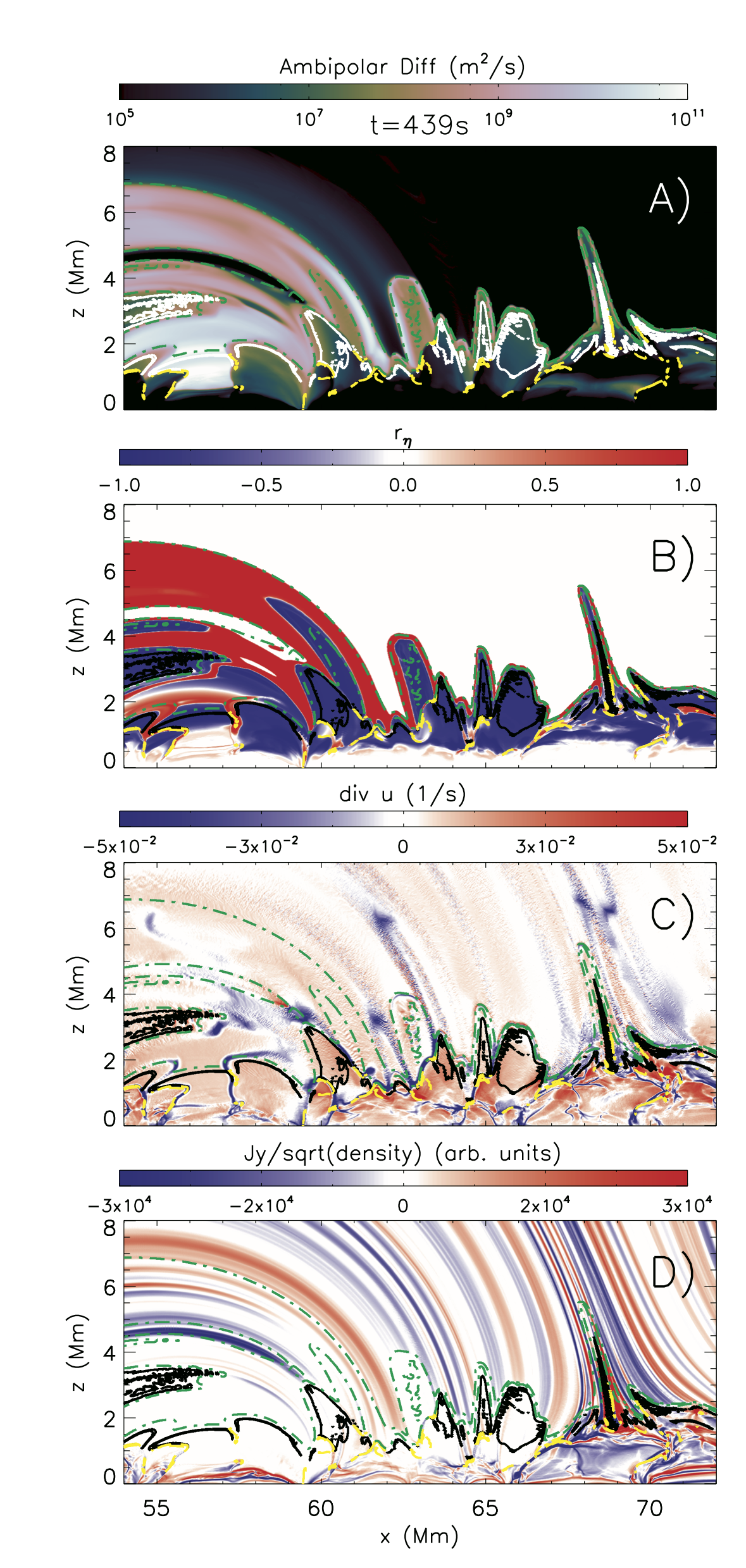}
	\caption{\label{fig:eta} Non-equilibrium ionization impacts largely the ambipolar diffusion.  {\jms For this figure we used only the \golneq\ simulation}. Panel A: ambipolar diffusion coefficient. Panel B: Difference in ambipolar diffusion between NEI and LTE ionization (Equation~\ref{eq:reta}). Panel C: divergence of the velocity. Panel D: ratio of current perpendicular to the simulated plane and density. Grid points within the solid white (panel A) or black (panels B-D) correspond to points within the contour labeled 1 in the bottom panel of Figure~\ref{fig:pdf}. The yellow dashed contour corresponds to the points within the contour labeled 2, and the green dash-dotted contour corresponds to the contour labeled 3 in Figure~\ref{fig:pdf}. An animation of this figure is available in the online Journal showing time evolution.}
\end{figure}
 
Figure~\ref{fig:eta} shows the important role of the ionization fraction under NEI conditions to the amplitude of ambipolar diffusion. Panel A of Figure~\ref{fig:eta} and associated animation shows the ambipolar diffusion coefficient of the \golneq\ simulation, panel B compares the ambipolar diffusion assuming NEI ionization ($\eta_\mathrm{A_{NEI}}$) with the diffusion assuming LTE ionization ($\eta_\mathrm{A_{LTE}}$) following the expression: 
\begin{eqnarray}
r_\eta = \frac{\eta_\mathrm{A_{NEI}}-\eta_\mathrm{A_{LTE}}}{\eta_\mathrm{A_{NEI}}+\eta_\mathrm{A_{LTE}}}. \label{eq:reta}
\end{eqnarray}
Red (blue) color means that ambipolar diffusion is underestimated (overestimated) under LTE conditions. Panel C shows the divergence of the velocity showing the location of strong compression in shocks (blue) and regions that undergo expansion (red). Finally, pane D shows the current perpendicular to the simulated plane which has been divided by the square root of the density in order to enhance the visibility of the currents at greater heights. 

\subsubsection{Cold expanding regions {\jms - using the \golneq\ simulation}}~\label{sec:ambcold}

We find that most of the fluid is highly ionized in the cold plasma forming during the early phases of the rise of cold bubbles, as well as in thin cold regions within type II spicules and low-lying loops (green, blue and red arrows in panels B and G in Figure~\ref{fig:denstg_zoom}).

In the low to mid chromosphere, the fraction of neutral hydrogen (H I, panel A in Figure~\ref{fig:ionf}) decreases at a height of $z\approx 1$~Mm, which also happens to always be at roughly at the same density of $10^{-5}$~kg~m$^{-3}$, see panel F in Figure~\ref{fig:denstg_zoom}). This is the location where the acoustic waves driven by convective motions at the photosphere become shocks due to the exponential density drop. Shocks appear (in blue in panel C of Figure~\ref{fig:eta}) as a result of strong compression and behind the shock fronts, i.e., in the wake, cold bubbles form as the plasma expands (red). Shocks passing through the middle-chromosphere ionize hydrogen (panel A in Figure~\ref{fig:ionf}). When non-equilibrium ionization is accounted for, shocks passing through the chromosphere ionize most of the hydrogen rapidly while the recombination timescales ($\sim 10^4$~s) are much longer than the hydrodynamic timescales (10-300~s). Consequently, the plasma stays ionized despite the drop in temperature in the cold bubbles (red in panel E in Figure~\ref{fig:ionf}). As a result of this, $\rho_{a\hat{I}}\nu_{a\hat{I}a'0}$ (see Equation~\ref{eq:diff}) is high and the coefficient of ambipolar diffusion is extremely low in NEI conditions (panel I in Figure~\ref{fig:denstg_zoom}). As a result, the effects of ambipolar diffusion in LTE are highly over-estimated in the cold bubbles (blue in panel B of Figure~\ref{fig:eta}).  It is only towards the end of the cold bubble evolution (a few minutes after the shock passes through) that the plasma starts to recombine and the effects of ambipolar diffusion increase in significance. 

In NEI, regions with strong expansion in the upper chromosphere and TR appear as red in panel C of Figure~\ref{fig:eta}. These regions are cold inside type II spicules and in low-lying loops but are relatively ionized in hydrogen, while being highly neutral in helium (panels B and C in Figure~\ref{fig:ionf}). As in the middle chromosphere, the long recombination time-scales 
as compared to the timescales of plasma expansion leads to a large percentage of ionized  hydrogen compared to what would be found in an atmosphere computed with the assumption of LTE (red in panel E in Figure~\ref{fig:ionf}). This leads to smaller effects of ambipolar diffusion in these strongly expanding regions than in the LTE case (blue regions in panel B of Figure~\ref{fig:eta}). 

These expanding regions (cold expanding bubbles and cold thin regions near the TR) are areas termed population 1 in the bottom panel of Figure~\ref{fig:pdf}, which will be described in further detail in Section~\ref{sec:adb}. Statistically speaking, this shows that the median of the coefficient of ambipolar diffusion $\eta_\mathrm{A}$ (panel F of Figure~\ref{fig:etapdf}) and $L_\mathrm{amb}$ (top panel of Figure~\ref{fig:lensclpdf}) are small for the plasma contained within the contour labeled 1 ($\eta_\mathrm{A} < 10^8$~m$^2$~s$^{-1}$). 

\subsubsection{Shocks, TR and low-lying loops {\jms - using the \golneq\ simulation}}~\label{sec:highamb} 

The opposite process happens in magneto-acoustic shocks or in the steep temperature rise of the TR  (green and yellow contours in Figure~\ref{fig:eta}). The number of neutrals in these regions is larger than expected under the assumption of LTE (blue in panel E of Figure~\ref{fig:ionf}). This is true for both hydrogen (panel A of Figure~\ref{fig:ionf}) in chromospheric shocks and for helium (panel B) in the upper chromosphere and TR. In these locations, the time between shocks passing is long enough that a large amount of plasma has managed to recombine. In addition, ionization does not happen immediately and furthermore the timescales of ambipolar heating are slightly shorter than the ionization timescales. 
All of these effects added together leads to lower ionization fractions in these regions in NEI than in LTE (blue in panel E in Figure~\ref{fig:ionf}). Thus, these locations have a larger ambipolar diffusion in NEI than in LTE (red in panel B of Figure~\ref{fig:eta}).  

The regions with the largest effects of ambipolar diffusion are found in low-lying TR or chromospheric magnetic loops at times long after shocks have passed through (e.g., around $x\approx [54-60]$~Mm in Figures~\ref{fig:ionf} and~\ref{fig:eta}). In these loops, the density decreases due to the draining of plasma along the loops towards the photosphere. Magneto-acoustic shocks passing through these highly inclined loops are less frequent than in other more vertical magnetic structures \citep[because the change in effective gravity leads to a lower acoustic cutoff frequency, e.g., ][]{Heggland:2007jt}.  Consequently, recombination timescales are short compared to the time that elapses between shocks propagating along these loops.
In this case neutrals become important, and, due to the low density, the ion-neutral collision frequency is low, resulting in an extremely high ambipolar diffusion. Additionally, these loops have more neutrals, especially in helium (panel B in Figure~\ref{fig:ionf}) than they would have assuming LTE (blue in panel E). The overabundance of neutrals is probably not due to the timescales involved, but rather that under these conditions the ionization state is better modeled by `coronal equilibrium' conditions, with collisional ionization being balanced by radiative recombination, which predicts more neutrals at a given temperature than what is found in LTE. This state of affairs is detailed for hydrogen in the top panels of Figure~4 of \citet{2016A&A...590A.124R}. Helium behaves in the same manner. A consequence is that the heating timescales due to ambipolar diffusion are much shorter than the ionization timescales. 
The ionization fraction, mostly in helium, in these regions can be severely overestimated when assuming LTE (panels B, C and E in Figure~\ref{fig:ionf} and Section~\ref{sec:high} for further details) and the ambipolar diffusion is larger in NEI than assuming LTE (red in panel B of Figure~\ref{fig:eta}).

In short, the effects of ambipolar diffusion are affected by the choice of assumption for the ionization balance. Assuming one scenario or the other can provide opposite roles for the ambipolar diffusion, i.e., in those locations where ambipolar diffusion is large in LTE (cold bubble or inside expanding spicules) it seems unimportant under conditions of NEI; whereas in those locations where the ambipolar diffusion is small in LTE (e.g., shock fronts), with NEI the ambipolar diffusion is large enough to play an important role as detailed below. 

\subsection{Impact of non-equilibrium ionization and ambipolar diffusion on the thermal evolution}~\label{sec:therm}

The previous section showed how NEI impacts ambipolar diffusion in the solar atmosphere. Now we detail how these changes in the ambipolar diffusion impact the thermodynamics of the plasma. 

\subsubsection{Adiabatic expansion}~\label{sec:adb}

Let us consider the physical processes that play a role in the cold regions that lie along type II spicules or in the cold bubbles formed by rarefaction behind magneto-acoustic shocks. These regions {\jms in the \golneq\ simulation} are visible within the black contours in panels B-D of Figure~\ref{fig:eta}. In the LTE models described by \citet{Martinez-Sykora:2017gol}, cold bubbles and spicules are heated via ambipolar heating the largest contributor. In those models the ambipolar diffusion increases drastically, reaching values that rapidly dissipate any current or Aflv\'enic wave passing through. The ambipolar diffusion becomes extremely large because the ion-neutral collision frequency and ionization degree diminish (panel D in Figure~\ref{fig:denstg_zoom} and panel E in Figure~\ref{fig:etapdf}) as plasma expansion causes the temperature and density to fall rapidly. Consequently, cold regions behind the shocks are vigorously heated by the ambipolar heating (panel E in Figure~\ref{fig:denstg_zoom}). However, this situation changes once NEI is taken into account (right column in Figure~\ref{fig:denstg_zoom}): we find that the cold bubbles are now much cooler \citep[see][but see our discussion for comparison with the literature]{Leenaarts:2011qy}. Similarly, elongated thin structures inside type II spicules can be much cooler in the \golneq\ simulation than in \gol\ (compare panel B and G in Figure~\ref{fig:denstg_zoom}).

Going back to the JPDF maps in Figure~\ref{fig:pdf}, we mentioned a very intriguing population {\jms in the \golneq\ simulation} indicated with a solid blue contour labeled 1 in the bottom panel. We added white or black contours in the maps in Figures~\ref{fig:eta} that correspond to the plasma within this population. There, one can see that this population corresponds to the cold expanding bubbles (see animated Figures~\ref{fig:eta}) as well as regions with strong expansion inside type II spicules. The slope of the contour labeled 1 in the bottom panel of Figure~\ref{fig:pdf} follows $\rho \propto T^{5/3}$. This corresponds to the adiabatic relation between density and temperature. Statistically speaking, in these regions, the plasma behaves adiabatically. This has two causes, 1) ambipolar diffusion is not large enough to dissipate any current in the NEI case as opposed to in LTE, so the entropy sources are minor (see below), 2) recombination timescales are larger than expansion timescales, at least for the early stages of the lifetime of the bubbles, so recombination does not change the plasma state significantly. 

In fact, in these regions the timescales of the energy losses due to expansion can be as low as $\frac{1}{\nabla \cdot {\bf u}} \approx 10$~s which is much smaller than recombination timescales ($\sim 10^4$~s). In other words, cooling from expanding regions in simulations assuming LTE has a ``reservoir" energy in the recombination energy of the ionized species. In contrast, in the simulation with non-equilibrium ionization/recombination (\golneq), the cooling due to expansion is not using the ``reservoir" energy in recombining the ionized species due to the rather large recombination timescales. Instead the temperature drops as dictated by the adiabatic expansion.

As mentioned before, in LTE {\jms (simulation \gol )}, the ambipolar diffusion increases abruptly  with expansion (panel D in Figure~\ref{fig:denstg_zoom} and associated animation) because of a decrease in ion-neutral collision frequency and ionization fraction (panel C). There, the ambipolar diffusion dissipates, or advects, almost any current or Alfv\'enic wave traveling through the cold bubbles (panels D-F). 

In contrast to this, when modeling NEI (simulation \golneq ), ambipolar diffusion is nonexistent in these regions (Section~\ref{sec:ambcold}) and the ambipolar heating therefore negligible (panel J in Figure~\ref{fig:denstg_zoom}). Figure~\ref{fig:hist} shows the histogram of the ambipolar diffusion (top) and of the ambipolar heating (bottom) for the 
\golneq\ simulation within the three selected regions: 1 (associated to the {\em adiabatic} expanding plasma, solid line), 2 (associated to the shocks, dashed line) and 3 (associated to plasma with low density, dotted line) shown in the bottom panel of Figure~\ref{fig:pdf}. Figure~\ref{fig:hist} confirms that the ambipolar diffusion is small and the ambipolar heating is negligible in the expanding cold bubbles and cold regions within the spicules (solid lines). 

{\jms In simulation \golneq ,} the Joule heating in cold bubbles is basically negligible, not only due to the rather low ambipolar diffusion, but also because most of the current and shocks are ahead of the cold bubble (panels C-D of Figure~\ref{fig:eta}). In spicules, the region that corresponds to the population 1 (white contours in panel A of Figure~\ref{fig:eta}) is highly confined along the spicule. Similarly, in these thin regions the Joule heating is smaller than in other locations within the spicule, since fewer and smaller currents are present. 

\subsubsection{Hot shocks and spicule strands {\jms in the NEI case}}~\label{sec:hot}

The second population of interest in this work has the opposite behavior. This population is indicated with the dashed yellow contour labeled 2 in the bottom panel of Figure~\ref{fig:pdf} {\jms for simulation \golneq }. The plasma comprising this population is localized mostly in the shock fronts or boundaries and/or footpoints of type II spicules (yellow contours in Figure~\ref{fig:eta} and corresponding animation).

\begin{figure}
	\includegraphics[width=0.49\textwidth]{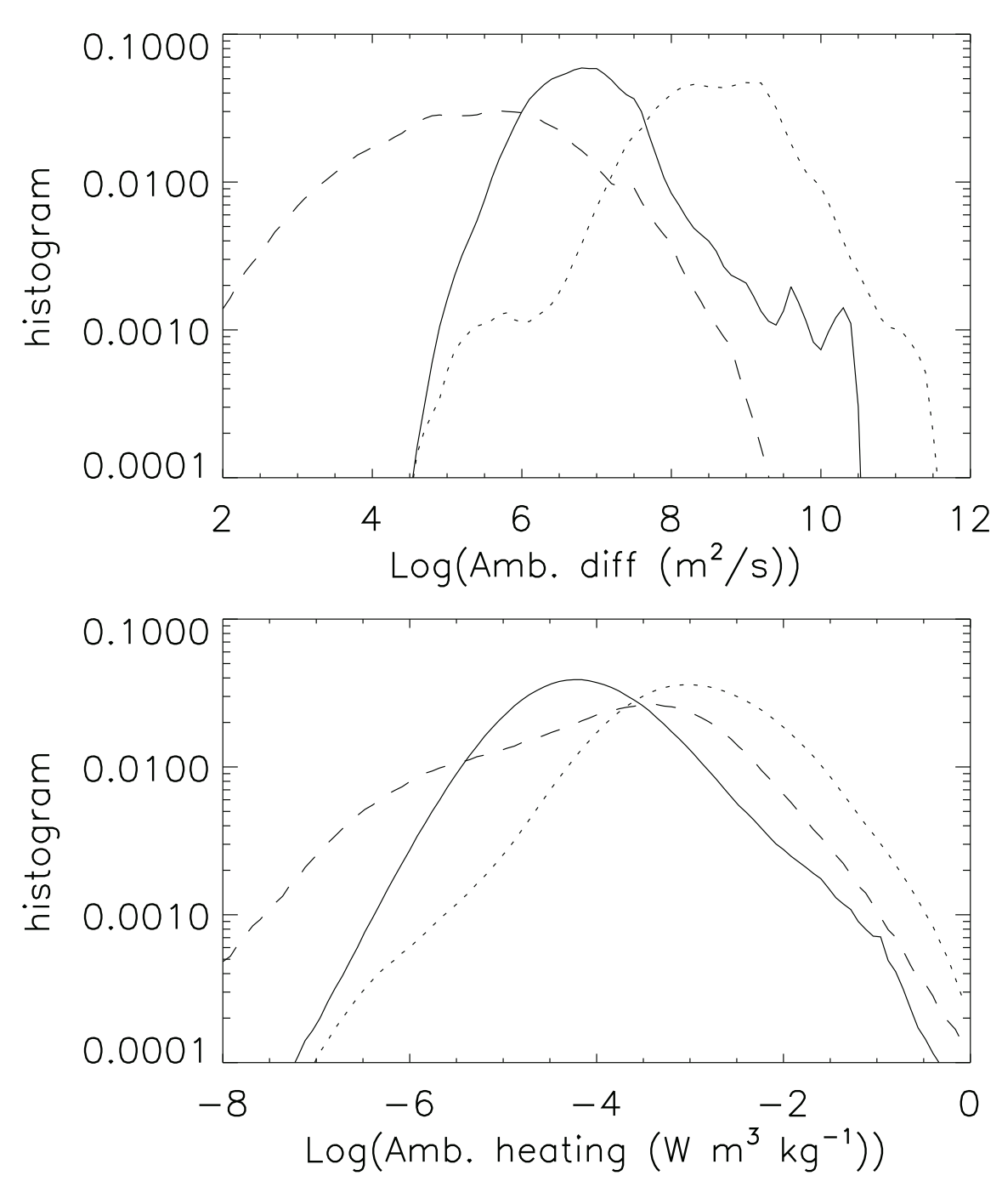}  
	\caption{\label{fig:hist} Histograms of the ambipolar diffusivity (top) and Joule heating through ambipolar diffusion (bottom) {\jms of the \golneq\ simulation} for the regions labeled 1 (associated with the {\em adiabatically} expanding plasma, solid line), 2 (associated with the shocks, dashed line) and 3 (associated with plasma with low density, dotted line)  in Figure~\ref{fig:pdf}. Spicules have more Joule heating from ambipolar diffusion than in the expanding cold bubbles; and low-lying loops have the largest ambipolar diffusion. }
\end{figure}

As mentioned above, in magneto-acoustic shocks and at sharp transitions with enough ambipolar diffusion and large currents, the ambipolar diffusion allows the dissipation of magnetic energy into heat. 
Therefore, ambipolar heating in the \golneq\ simulation contributes to heating these locations (panel J in  Figure~\ref{fig:denstg_zoom} and corresponding animation). In fact, the ambipolar heating  (dashed line in bottom panel of Figure~\ref{fig:hist}) is up to an order of magnitude larger than inside the expanding cold bubbles (solid). The heat in shocks and those locations in spicules  with large ambipolar diffusion comes not only from  ambipolar heating, but also artificial diffusion, viscous heating and compression. The timescales of all these heating processes can reach a similar order of magnitude ($\sim 1$~min).  

It is interesting to consider what this rapid heating in spicules means for a long-standing open question about the rapid apparent motions associated with spicules visible in TR lines. 
There are conflicting observations between type II spicules (typically observed spectrally with lines shifts showing line-of-sight velocities of up to $\sim 100$~km~s$^{-1}$), and apparent fast motions (the so-called network jets that show apparent speeds in the plane of the sky of a few hundred of km~s$^{-1}$) \citep{Beck:2013A&A...556A.127B,Tian:2014fp,Rouppe-van-der-Voort:2015fr}. \citet{DePontieu:2017net} have shown that network ``jets" do not necessarily involve very high plasma flows but can be caused by rapidly propagating heating fronts associated with ambipolar dissipation of currents traveling along type II spicules. The quoted work assumed LTE. In the current work we relax the LTE assumption and investigate NEI conditions. The NEI calculations do not fundamentally change this picture. We again find that the extremely large currents in spicules (see bottom panel of Figure~\ref{fig:eta}) are dissipated via ambipolar diffusion. The extremely large currents explain how this population (dashed line vs solid line in Figure~\ref{fig:hist}) can be subject to much larger ambipolar heating than what is found in cold expanding bubbles, despite the relatively smaller ambipolar diffusion in spicules. For instance, the example shown in panel J in Figure~\ref{fig:denstg_zoom} reveals large heating occurring along the boundaries of (and inside of) the spicules. In other words, the NEI effects mostly affect the spatial distribution and (local) amount of heating, but do not not impact how fast a propagating current or Alfv\'en wave is propagating or dissipated. As in shock fronts, artificial diffusion, viscous heating and compression also play an important role. In addition, spicules have  temperature gradients high enough that thermal conduction contributes substantially to their thermal properties.  The importance of each of these source terms varies within the spicules and evolution.

Due to the relatively large ionization timescales (a few tens of seconds) in these regions of the \golneq\ simulation, heating goes directly to increasing the plasma temperature instead of ionizing the plasma. Consequently, magneto-acoustic shocks are much hotter in the \golneq\ simulation. In  spicules we find both: 
1) cold regions with low ambipolar diffusion, low currents, large expansion, and large  recombination time-scales, and
2) hot regions with high ambipolar diffusion and large currents. These regions experience strong a temperature increase due to long ionization timescales. 
As a result of these two properties, spicules in \golneq\ have large temperature variations within them. Note that, roughly when the spicule has fully formed, at its boundaries, the ambipolar diffusion can get as high as $10^{10}$ m$^2$ s$^{-1}$ (panel A in animated Figure~\ref{fig:eta}). The large density and temperature variations within spicules may explain the apparently conflicting observations of temperatures and speeds in spicules \citep[compare][]{Beck:2013A&A...556A.127B,Tian:2014fp,Rouppe-van-der-Voort:2015fr}. 

Another important aspect is that type II spicules seem to impact the associated coronal loops \citep{Henriques:2016ApJ...820..124H,De-Pontieu:2017pcd}. Due to NEI effects, the amount of current (which in our models causes heating of coronal plasma) that pierces into the corona is larger than in the LTE case and its role in the corona seems to be larger than what was previously reported by  \citet{Martinez-Sykora:2017rb}. We find that the corona has on average 10\% more current density in the \golneq\ simulation than in the \gol\ simulation. A proper calculation of synthetic observables from this simulation and comparison with observations is required for further studies on the apparent motions and impact of spicules on the corona.

\subsubsection{Low density low-lying loops with large ambipolar diffusion}~\label{sec:high}

As mentioned above, {\jms in NEI,} low-lying loops with either chromospheric or transition region temperatures show the largest ambipolar diffusion ($\eta_\mathrm{A}>10^{8}$~m$^2$~s$^{-1}$, Section~\ref{sec:highamb}).  These regions are relatively low density ($10^{-12} < \rho<10^{-9}$~kg~m$^{-3}$) and with temperatures $T<10^5$~K (labeled as region 3 and shown with dash-dotted green contours in Figures~\ref{fig:pdf}, \ref{fig:eta} and with dotted lines in Figure~\ref{fig:hist}). 

Since the ambipolar diffusion is larger in NEI than assuming LTE in low-lying loops (Section~\ref{sec:highamb}), the heating is also larger in NEI than in LTE. In  NEI, heating from the ambipolar diffusion is more than an order of magnitude larger in these loops than in the other features described in the previous sections (dotted line in bottom panel of Figure~\ref{fig:hist}). This is because the ambipolar diffusion coefficient is several orders of magnitude larger in these loops (dotted line in top panel of Figure~\ref{fig:hist}) than in the other features (dashed or solid lines). Consequently, almost any Alfv\'enic wave or current is dissipated there. 

These low-lying loops share similarities with two different types of loops that are observed in the solar atmosphere: 1. loops that appear to outline the canopy and connect plage regions of opposite polarity in active regions, typically clearly visible in chromospheric (e.g., H$\alpha$) or TR (e.g., \ion{Si}{4} 1400\AA) diagnostics; 2. so-called Unresolved Fine Structure \citep[UFS,][]{Feldman:1983so,Hansteen17102014}.

Our results suggest that ``canopy loops" may be composed of plasma with a large range of temperatures {\jms due to NEI effects, ``adiabatic" expansions, and ambipolar heating}, with a mix of hot and cool strands right next to each other. In our simulations, these low-lying loops initially are filled with cool plasma due to shocks passing through from one footpoint all the way to the other footpoint (see animated Figures~\ref{fig:denstg_zoom}, \ref{fig:ionf}, and \ref{fig:eta}). After the cold loop has formed, the fluid recombines, ambipolar diffusion increases and the loop is heated up to TR temperatures via current dissipation from the ambipolar diffusion. This could explain the presence of chromospheric and TR loop-like structures that both appear to outline the canopy and occur in close proximity of one another.

Similarly, observations suggest that UFS loops seems to be formed by sporadic heating instead of cooling from hot loops \citep{Hansteen17102014,Brooks:2016ao}. 3D radiative MHD models (excluding ambipolar diffusion) show that the cycling of TR plasma starts from being first rapidly heated from chromospheric temperatures to later drain and cool down in small magnetic loops \citep{Guerreiro:2013ApJ...769...47G}. 
Our model suggests that the ambipolar diffusion may play an important role in the thermal evolution of these low-lying loops: The ambipolar diffusion dissipates Alfv\'enic waves or currents into thermal energy. Since the ambipolar heating timescales are shorter or comparable to ionization timescales, most of the heating goes into heating the plasma instead of ionizing the plasma. In addition, the ambipolar diffusion is important all the way up to TR temperatures, even above helium ionization temperatures in LTE, owing to the large role of NEI as shown in Figures~\ref{fig:denstg_zoom}, \ref{fig:etapdf} and \ref{fig:ionf}.

\section{Discussion and Conclusions}~\label{sec:dis}

We have performed a 2.5D radiative-MHD simulation that self-consistently includes hydrogen and helium in non-equilibrium ionization and ion-neutral interaction effects (ambipolar diffusion and the Hall term). We  have compared this model with a previously-existing model which included only the ion-neutral interaction effects. Our results reveal that combining hydrogen and helium in NEI and ion-neutral interaction effects leads to substantial thermal differences compared to the model with LTE ionization. Ambipolar diffusion is influential in the solar chromosphere, but plays different and opposing roles depending on whether one assumes LTE or NEI when computing populations. In short, in NEI, the ambipolar diffusion and its
Joule heating play a role in the thermodynamics in shock fronts, low-lying chromospheric and TR loops, and hot regions within type II spicules. In contrast, in LTE, the role of the ambipolar diffusion and associated Joule heating is negligible in shock fronts and important in post-shock plasma and cold regions within type II spicules. 

The thermodynamic evolution of plasma in wakes produced by the rarefaction behind the magneto-acoustic shocks in the chromosphere resembles previous 2D radiative MHD models in revealing the large role of the NEI effects \citep{Leenaarts:2007sf}. During the first tens of seconds of the wake formation, the expansion is quasi-adiabatic where the entropy sources are minor.  However, 
those previous simulations do not reveal the adiabatic expansion of population 1, most likely, because plasma reached very rapidly the minimum temperature allowed by the imposed ad hoc heating \citep{Leenaarts:2007sf,Leenaarts:2011qy}. Most of the chromospheric plasma in their simulation was much cooler than in \golneq\ most likely due to a very weak magnetic field and lack of ambipolar diffusion in the former. At that temperature, an {\it ad hoc} heating term is added to avoid plasma temperatures outside the validity range of the equation of state. 

Type II spicules have large temperature variations owing to non-equilibrium properties of hydrogen and helium and ion-neutral interaction effects. 
Some plasma elements within the spicules become very cold due to a quasi-adiabatic expansion where the entropy sources are minor. 
Other plasma elements get heated due to strong currents traveling at Alfv\'en speed and dissipated via ambipolar and artificial diffusion \citep{Tian:2014fp, Narang:2016db,DePontieu:2017net}. In both scenarios plasma is in NEI, and heating and cooling mostly change the temperature instead of ionizing or recombining the plasma. The heating associated with rapidly propagating currents and the complex temperature distribution within spicules may explain apparent discrepancies between various observations of temperature, velocities, and apparent speeds, by, e.g., \citet{Rouppe-van-der-Voort:2015fr}, \citet{Beck:2013A&A...556A.127B}, and \citet{Tian:2014fp}.

Our \golneq\ model suggest that low-lying loops outlining the canopy (e.g., in H$\alpha$ and TR lines) may be composed of multi-thermal threads that are formed by shocks passing through these loops. These shocks fill the loops with cold plasma, and due to the slow thermal evolution compared to recombination timescales, ambipolar diffusion becomes important and heats the low-lying loops. Ambipolar diffusion becomes extremely large and leads to strong heating up to TR temperatures. Heating timescales are thus shorter than ionization timescales and the heating goes mostly towards increasing the temperature. We also speculate that heating in UFS loops may be caused by a similar mechanism. 

In summary, heating from either the ambipolar or artificial diffusion, compression, or thermal conduction does not go directly towards ionizing the plasma, but instead it increases the temperature. Consequently, shocks, low-lying loops and regions with large current in type II spicules are heated to greater temperatures than in the LTE models. Similarly, when quasi-adiabatic expansion takes place, the cooling does not use the reservoir energy from the recombination, instead the plasma cools down. As a result of this, cold bubbles with non-efficient entropy sources have an adiabatic relation between temperature and density, in contrast to what occurs in models based on LTE. 

It is critical to properly treat the physical processes and the thermodynamics in the numerical models in order to correctly interpret the observations. Our results show that NEI and ambipolar diffusion play a significant role in the thermodynamics. These physical processes potentially have a large impact on opacities and chromospheric spectral profiles. Forward modeling of chromospheric lines from these models may thus shed light on the interpretation of complex chromospheric observations. 

Our simulations are limited to 2.5 dimensions. In three dimension it is expected that shocks and cold bubbles, magnetic reconnection and currents may change quantitatively compared to a 2.5D scenario \citep[e.g., ][]{Martinez-Sykora:2019dyn}. In addition, in 3D the convective motions build up more current in the atmosphere than in 2D. Despite these differences, we expect that our results are qualitatively robust thanks to the self-consistent implementation of the many processes included in our models. 

\section{Acknowledgments}

We gratefully acknowledge support by NASA grants NNX16AG90G, NNX17AD33G, 80NSSC18K1285 and contract NNG09FA40C (IRIS), NSF grant AST1714955. 
JL was supported by a grant (2016.0019) of the Knut and Alice Wallenberg foundation. The simulations have been run on clusters from the Notur project, and the Pleiades cluster through the computing project s1061, s1630, s1980, and s2053 from the High End Computing (HEC) division of NASA. This study has been discussed within the activities of team 399 ``Studying magnetic-field-regulated heating in the solar chromosphere" at the International Space Science Institute (ISSI) in Switzerland. To analyze the data we have used IDL. 
This research is also supported by the Research Council of Norway through 
its Centres of Excellence scheme, project number 262622, and through 
grants of computing time from the Programme for Supercomputing.

\bibliographystyle{aa}
\bibliography{aamnemonic,collectionbib}

\end{document}